\documentclass[]{article}

\newcommand{\ga}{\alpha} \newcommand{\gb}{\beta}
\newcommand{\gc}{\gamma}

\newcommand{\sbs}{\subseteq} 
 \newcommand{\lto}{\rightarrow}
\newcommand{\sat}{{\models}}

\newcommand{\vd}{\vdash}

\newcommand{\imp}{\rightarrow}

\newcommand{\qed}{\vrule height5pt width3pt depth0pt}
\newcommand{\pev}{\makebox[1.3em]{$\rule{0.1mm}{2.2mm}\hspace{-1.3mm}\prec$}}
\newcommand{\notpev}{\makebox[1.3em]{$\rule{0.1mm}{2.2mm}\hspace{-1.3mm}\not\prec$}}
\newcommand{\ev}{\makebox[1.3em]{$\rule{0.1mm}{2.2mm}\hspace{-1.2mm}\sim$}}
\newcommand{\notev}{\makebox[1.3em]{$\rule{0.1mm}{2.2mm}\hspace{-1.2mm}\not\sim$}}

\title{Relevance as Deduction: A Logical View of Information
Retrieval
\\ \normalsize (Abstract)
}
\author{\normalsize
\begin{tabular}{cc}
 Gianni Amati & Konstantinos Georgatos \\
Fondazione Ugo Bordoni & Dipartimento di
Informatica e Sistemistica\\
Via B. Castiglione 59, 00142 Roma, Italy & Universit\`a di Roma ``La
Sapienza''\\
 e-mail: {\tt gba@fub.it} & Via Salaria 113, 00198 Roma, Italy
\\ & e-mail: {\tt geo@dis.uniroma1.it}
\end{tabular}
 }

\date{July 19, 1996}

\begin{document}
\maketitle

The problem of Information Retrieval is, given a set of documents $D$ and a
query $q$, providing an algorithm for retrieving all documents in $D$ relevant
to $q$. However, retrieval  should depend and be updated whenever the
user is able to provide as an input  a preferred set of relevant documents;
this process
is known as {\em relevance feedback}. Recent work in IR has been paying
great attention
to models which employ a logical approach;
the advantage being  that one can have a simple computable
characterization of
retrieval on the basis of a pure logical analysis of retrieval. Most of the
logical
models make use of probabilities or similar belief functions in order to
introduce the
inductive component whereby  uncertainty is treated. Their general paradigm is
the following: {\em find the nature of conditional $d\imp q$ and then define a
probability on
the top of it\/}. We just reverse this point of view;  {\em first use the
numerical information,  frequencies or probabilities, then define your own
logical consequence\/}.
More generally, we claim that retrieval is a form of deduction. We
introduce a simple but
powerful logical framework of relevance feedback, derived from the well
founded area of
nonmonotonic logic.  This description can help us evaluate, describe and
compare  from a
theoretical point of view previous approaches based on conditionals or
probabilities.

The first difficulty one encounters towards a logical approach to Information
Retrieval is  how one should view the algorithm which returns a set of
relevant documents. Our proposal is to see such an algorithm as a {\em
proof}. However, this proof is not classical because, as it is widely
recognized,
$q\lto d$, where $q$ is a query and $d$ a document relevant to $q$, is not
conveyed by
the material implication (see~\cite{RIJ86},\cite{RIJ89}). Also, as the set of
data which this proof is based  is subject to updating, nonmotonicity arises.
Since there is no need of nesting implications, implication and deduction can be
identified and therefore  our starting point is to axiomatize a deduction $q\ev
d$. The consequence relation
$\ev$ will reflect the
properties that relevance satisfies. A similar approach where nonmonotonic
consequence relations are used to axiomatize a notion of {\em aboutness\/}
appears in \cite{BH95}. We shall then show how
$\ev$ changes with relevance feedback.

A set of rules for the relevance relation $\ev$ appears in
Table~\ref{table:rules}. This set of rules apart from Rational Monotonicity
comprises the system
$P$ (see~\cite{KLM90}) of {\em preferential inference\/}. Preferential inference along
with the rule of Rational Monotonicity comprises the system $R$ of {\em
rational inference\/} (see~\cite{LM92}).  We propose preferential and rational
inference as the simplest and strongest systems axiomatizing the notion of
relevance.

\begin{table}[t]
\begin{center}
\begin{minipage}{4.5in}
%\begin{minipage}{5in}  % for fullpage
{\small
\begin{tabular}{cl}

$\displaystyle \frac{\ga \vd \gb}{\ga\ev\gb}$ & {\footnotesize
(Supraclassicality)}
\vspace{.1in} \\

$\displaystyle \frac{\vd\ga\equiv\gb \qquad \ga\ev\gc}{\gb\ev\gc}$ & {\footnotesize
(Left Logical Equivalence)} \vspace{.1in}\\

$\displaystyle \frac{\ga\ev\gb \qquad \gb\vd\gc}{\ga\ev\gc}$ &
{\footnotesize(Right Weakening)} \vspace{.1in}\\

$\displaystyle \frac{\ga\ev\gb \qquad \ga\ev\gc}{\ga\ev\gb\land\gc}$ &
{\footnotesize(And)} \vspace{.1in}\\

$\displaystyle \frac{\ga\ev\gb \qquad
\ga\land\gb\ev\gc}{\ga\ev\gc}$ & {\footnotesize(Cut)} \vspace{.1in}\\

$\displaystyle \frac{\ga\ev\gb \qquad
\ga\ev\gc}{\ga\land\gb\ev\gc}$ & {\footnotesize(Cautious
Monotonicity)}\vspace{.1in}\\

$\displaystyle \frac{\ga_0\ev\ga_1\quad\cdots\quad \ga_{n-1}\ev\ga_n\quad
\ga_n\ev\ga_0}{\ga_0\ev\ga_n}$ & {\footnotesize(Loop)}\vspace{.1in}\\

$\displaystyle \frac{\ga\ev\gc \qquad
\gb\ev\gc}{\ga\lor\gb\ev\gc}$ & {\footnotesize(Or)} \vspace{.1in}\\

$\displaystyle \frac{\ga\notev\neg\gb \qquad
\ga\ev\gc}{\ga\land\gb\ev\gc}$ & {\footnotesize(Rational Monotonicity)}
\end{tabular} }
\end{minipage}

\end{center}
\bigskip

\caption{\label{table:rules} Rules for Nonmonotonic Inference}
\end{table}

However, it is well known that designing a nonmonotonic system is far from
trivial. Moreover, nonmonotonicity does not correspond directly to statistical
data as statistical operations are in general continuous or monotonic. A
solution to this problem is given by a recent characterization of the systems
of preferential and rational inference through systems of {\em monotonic\/}
consequence relations (\cite{KG96f}) called {\em priority relations\/}.  A
presentation of priority relations follows. Our language of terms will contain
only conjunction ($\land$). The reason we do not employ a notion of negation is
that there is no notion of negation which reflects our intuition in IR. The
standard way to look at negation is ruled by the
{\em closed world assumption\/}: if a term does not occur in a document then
its negation
occurs. However this makes documents possible worlds and deduction would turn
out classical in our model.  Hence  negation will be replaced by another
construction (see below). Documents will be identified with the conjunction of
all terms that occur in them. So, if
$\{t_1,\ldots,t_n\}$ are all documents occurring in a document $d$, then
$d=t_1\land\ldots\land t_n$. Our model is based on a recent characterization of
nonmonotonicity by the second author through a family of relations among
formulas called
{\em priority relations\/}. A priority relation satisfies\\
\begin{tabular}{rll} 1. &  $ t_1\pev t_1$ & (Reflexivity) \\
2. &  $ t_1\sat t_2$ and $ t_2\pev  t_3 $ implies $  t_1 \pev  t_3 $ &
      (Monotonicity) \\
3. & $ t_1\sat t_2$ and $ t_2\sat t_1$ implies $ t_3\pev t_1 $
iff
$ t_3\pev t_2$,& (Logical Equivalence)
\end{tabular}\\
where $\sat$ is classical. If a priority relation satisfies, in addition,\\
\begin{tabular}{rll}
4. & $ t_3\pev t_1$ and $ t_3\pev t_2$ implies
$ t_3\pev t_1\land t_2$,& (Right Conjunction)\\
5. & $ t_1\pev t_2$ and $ t_2\pev t_3$ implies
$ t_1\pev t_3$,& (Transitivity)
\end{tabular}\\
it will be called {\em preferential ordering\/}.
If a preferential ordering satisfies, in addition,\\
\begin{tabular}{rll}
6. & $ t_1\pev t_2$ or $ t_2\pev t_1$ & (Connectivity)
\end{tabular}\\
it is called {\em rational ordering\/}.

We now have the following theorem (\cite{KG96e},\cite{KG95e}): {\em Preferential
and Rational inference relations are generated by preferential and rational
orderings, respectively, through a maxiconsistent inference scheme. Moreover,
this correspondence is bijective.}  This is a powerful machinery
for handling nonmonotonic deductions. Once we
 build a
preference (preferential or rational) ordering, we are able to define and
compute the
associated nonmonotonic consequence relation. Turning conjunction to union of
sets of terms the defining conditions of priority relations become:
\\
\begin{tabular}{rll} 1. &  $ U\pev U$ & (Reflexivity) \\
2. &  $ U\sbs V$ and $ U\pev  W $ implies $  V \pev W $ & (Monotonicity) \\
3. & $ U\pev V$ and $ U\pev W$ implies
$ U\pev V\cup W$,& (Right Union)\\
4. & $ U\pev V$ and  $V\pev W$ implies
$ U\pev W$,& (Transitivity)\\
5. & $ U\pev V$ or $ V\pev U$ & (Connectivity)
\end{tabular}

One can think of these {\em monotonic\/} consequence relations as  {\em
aboutness} relations. That aboutness should be represented as a monotonic
consequence relation appears already in (\cite{HD95}) along with a mix of
nonmonotonic rules. Moreover, the above duality theorem translates to the fact
that relevance and aboutness are really dual notions. We shall now show how one
can easily generate a preferential ordering from frequency information. We show
how the user's relevance defines a nonmonotonic consequence relation which is
represented by a preferential ordering. In IR, a way to define the informative
content of terms is given by the frequencies in the document collection (as in
the Robertson Sparck Jones formula): this is enough for us to carry on
deductions! A way of constructing a preferential ordering among terms is to
divide the set of  documents  that the user distinguishes into two subsets,
the  positive set ($D^+$) and negative set ($D^-$). Then the set of positive
and negative examples will use frequencies to construct two rational orderings.
Then these orderings will be combined in a preferential ordering which reflects
our intuitions for relevance. Then a document
$d=t'_1\land\ldots\land t'_m$ will be relevant to query
$q=t_1\land\ldots\land t_n$ just in case $t_1\land\ldots\land t_n \pev
t'_1\land\ldots\land t'_m$.
The construction is the following. Denote with $D^+_t$ the set of positive
relevant documents where $t$ occurs, similarly for the negative case. The
frequency of $t$ relative to $D^+$ is $|D^+_t|$. Now define the following
orderings
$$t_1 \pev_p t_2 \quad \hbox{iff} \quad  | D^+_{t_1} | \leq | D^+_{t_2} |.$$
and
$$t_2 \pev_n t_1 \quad \hbox{iff} \quad  | D^-_{t_1} | \leq |  D^-_{t_2} |.$$
It can be shown that $\pev_p$ and $\pev_n$ are rational orderings.
Now let
$$t_1 \pev t_2 \quad \hbox{iff} \quad  t_1\pev_p t_2\ \hbox{and}\ t_1\pev_n t_2.$$
The consequence relation $\pev$ is a preferential ordering. It is now clear that
if we perform an update of the sets of positive and negative relevant documents
with relevance feedback then $\pev$ changes, too. In particular, this change is
{\em  nonmonotonic\/}.

Consider the following example. The following matrix is the matrix of occurrences
between documents $d_1,d_2,d_3,d_4$ and terms $t_1,t_2,t_3,t_4$.

\begin{center}

\begin{tabular}{r|c|c|c|c|l}
      & $t_1$ & $t_2$ & $t_3$ & $t_4$  &    \\ \hline
$d_1$ & 0     & 1     & 0     & 1      & $+$\\
$d_2$ & 0     & 1     & 1     & 0      & $+$\\
$d_3$ & 0     & 0     & 1     & 1      & $-$\\
$d_4$ & 1     & 0     & 0     & 1      & $-$
\end{tabular}

\end{center}

We have $t_1\pev_p \{ t_3,t_4 \} \pev_p t_2$ and $t_4 \pev_n \{t_1, t_3\} \pev_n
t_2$. Thus $t_1\pev t_3 \pev t_2$ and $t_4 \pev t_2$. For example, $t_3\land t_4\pev
d_1$ and $t_3\land t_4\pev d_3$, that is the query $t_3\land t_4$ returns $d_1$
and $d_3$ as relevant documents, but not $d_2$ and $d_4$.
This provides a model for computing  the relevance of documents with
respect a query
on the basis of some evidence.

  Consider now a {\em new} document $d_5=t_1\land t_2$. We
have that
$d_5$ is not totally ranked among the rest of the documents: we can decide
just $d_5\pev
d_2$ and
$d_4\pev d_5$. Let $U$ be the number of undecided documents, $D^+_p$ be the
set of
positive documents $d$ such that $d\pev d_5$, $D^-_n$ be the set of
negative documents  $d$ such that $d_5\pev d$,  $D^+_n$ be the set of
positive documents  $d$ such that $d_5\pev d$, and $D^-_p$ be the set of
negative documents  $d$ such that $d\pev d_5$.  If we want to create such a
linear ordering among documents we can use the the Robertson Spark Jones'
formula  and relate it to the number of incomparable documents. Then
$$ \frac{N-U}{N}\log\left( \frac{D^+_p \times D^-_n}{D^-_p \times
D^+_n}\right).$$
can be used as a decision rule. In the above example,  we get a negative value
for
$d_5$ which
is then chosen as irrelevant.

A point whose importance  should be stressed  is that generation of a
nonmonotonic consequence relation (relevance) through priority relations
depends on a notion of negation. Only in presence of negation we are able to
define maxiconsistent inference (see~\cite{KG96f}). We shall now show how a
recent approach to IR (\cite{ARU96}) using expected utility functions defines a
natural notion of negation. Suppose that the user supplies two sets of documents
representing positive and negative examples. Through theses sets and the
frequencies of terms appearing in those sets one  can construct for every
term a contingency matrix. Using now {\em entropy\/} $H$ and Hintikka's content
$C$, one can define a {\em weighting} function from the set of terms to the
interval
$[-1,1]$. Let $w$ be the weighting function. Set $w(\neg t)=-w(t)$ and $w(t\land
t')=\min(w(t),w(t'))$. Let $$r(t)=\left\{ \begin{array}{ll}
w(t) & \mbox{if $w(t) >0$}\\
0    &  \mbox{otherwise}
\end{array}
\right.$$
The function $r$ {\em ranks\/} terms according to their utility. Now the
following relation
$$t_1\pev t_2 \qquad\mbox{iff}\qquad r(t_1)\leq r(t_2)$$
is a rational ordering, and the inference
$$t_1\ev t_2 \quad\mbox{iff}\quad t_1\land t_3 \vd t_2,\ \mbox{for some}\
t_3\notpev\neg t_1$$
is a rational {\em nonmonotonic\/} inference. The above relation can be
readily extend to all boolean combinations of terms. However, we consider only
conjunctions  of positive terms. Then the rational inference can be
effectively checked through the following equivalent definition ($t_1$ and
$t_2$ are sets of terms)
\begin{center}
\begin{tabular}{rcl}
$t_1\ev t_2$ & iff & either $t_2 \sbs t_1$,\\
             &     & or $t_2 - t_1 \notpev \neg t_1$.
\end{tabular}
\end{center}

As an example consider three terms $t_1$, $t_2$ and $t_3$ and their contingency
matrices based on a set of $300$  relevant documents and a set of $700$
irrelevant documents:
\begin{center}
\begin{tabular}{r|c|c|}
                 & $Rel$ & $\overline{Rel}$           \\ \cline{2-3}
           $t_1$ & 0     & 300                     \\ \cline{2-3}
$\overline{t_1}$ & 626      &  74                         \\ \cline{2-3}

\end{tabular}\qquad
\begin{tabular}{r|c|c|}
           & $Rel$ & $\overline{Rel}$           \\ \cline{2-3}
      $t_2$ &  24    &   276                \\ \cline{2-3}
$\overline{t_2}$ & 676     &  24                \\ \cline{2-3}
\end{tabular}\qquad
\begin{tabular}{r|c|c|}
           & $Rel$ & $\overline{Rel}$           \\ \cline{2-3}
      $t_3$ & 290    &  10                \\ \cline{2-3}
$\overline{t_3}$ & 47    &   653              \\ \cline{2-3}
\end{tabular}
\end{center}

The weighting function now gives the following
\begin{center}
\begin{tabular}{rcl}
$w(t_1)$ & $=$ & $-.802$ \\
$w(t_2)$ & $=$ & $-.885$  \\
$w(t_3)$ & $=$ & $.845$   \\
$w(t_1\land t_2)$ & $=$ & $-.885$
\end{tabular}
\qquad
\begin{tabular}{rcl}
 $r(t_1)$ & $=$ & $0$  \\
 $r(t_2)$ & $=$ & $0$ \\
$r(t_3)$ & $=$ & $.845$  \\
 $r(t_1\land t_2)$ & $=$ & $0$
\end{tabular}
\end{center}

We now have
$t_1\ev t_3$, since $.845\not\leq .802$, but $t_1 \land t_2\notev t_3$, since
$.845\leq.885$. A table of all possible derivations between terms appears
below:
\begin{center}
\begin{tabular}{r|c|c|c|c|c|c|}
$q \ev d $  & $t_1$ & $t_2$ & $t_3$ & $t_1\land t_2 $ & $t_1\land t_3$ & $t_2
\land t_3 $ \\ \hline
$t_1$            & Y & N & Y & N & Y & N \\ \hline
$t_2$            & N & Y & N & N & N & N \\ \hline
$t_3$            & N & N & Y & N & N & N \\ \hline
$t_1 \land t_2 $ & Y & Y & N & Y & N & N \\ \hline
$t_1 \land t_3 $ & Y & N & Y & N & Y & N \\ \hline
$t_2 \land t_3 $ & N & Y & Y & N & N & Y \\ \hline
\end{tabular}
\end{center}

This logical framework is just an initial suggestion for future research,
which shows
that nonmonotonicity, induction and IR can have many common features which are
worthwhile to explore.


\begin{thebibliography}{AvRU96}

\bibitem[AvRU96]{ARU96}
G.~Amati, C.~J. van Rijsbergen, and F.~Ubaldini.
\newblock The maximum expected utility principle and information retrieval.
\newblock In D.L. Dowe, K.B. Korb, and J.J. Oliver, editors, {\em Proceedings
  of the Conference on Information, Statistics and Induction in Science}, pages
  129--140, Singapore, 1996. World Scientific Publishing Co.

\bibitem[BH95]{BH95}
P.~D. Bruza and T.~W.~C. Huibers.
\newblock How nonmonotonic is aboutness?
\newblock Technical Report CS-1995-09, Department of Computer Science, Utrecht
  University, The Netherlands, March 1995.

\bibitem[Geo]{KG95e}
Konstantinos Georgatos.
\newblock Preferential orderings.
\newblock Typed Manuscript. July 1995.

\bibitem[Geo96a]{KG96e}
Konstantinos Georgatos.
\newblock Ordering-based representations of rational inference.
\newblock In Jos\'e~J\'ulio Alferes, Lu\'is~Moniz Pereira, and Ewa Orlowska,
  editors, {\em Logics in Artificial Intelligence (JELIA '96)}, number 1126 in
  Lecture Notes in Artificial Intelligence, pages 176--191, Berlin, 1996.
  Springer-Verlag.

\bibitem[Geo96b]{KG96f}
Konstantinos Georgatos.
\newblock Expectation relations: A uniform approach to nonmonotonicity.
\newblock Working Paper 35--96, Dipartimento di Informatica e Sistemistica,
  Universit\'a di Roma ``La Sapienza'', November 1996.

\bibitem[HD95]{HD95}
T.~W.~C. Huibers and N.~Denos.
\newblock A qualitative ranking method for logical information retrieval
  models.
\newblock Technical Report RAP95-005, Groupe MRIM of the LGI in Grenoble,
  France, August 1995.

\bibitem[KLM90]{KLM90}
S.~Kraus, Daniel Lehmann, and Menachem Magidor.
\newblock Nonmonotonic reasoning, preferential models and cumulative logics.
\newblock {\em Artificial Intelligence}, 44:167--207, 1990.

\bibitem[LM92]{LM92}
Daniel Lehmann and Menachem Magidor.
\newblock What does a conditional knowledge base entail?
\newblock {\em Artificial Intelligence}, 55:1--60, 1992.

\bibitem[vR86]{RIJ86}
C.~J. van Rijsbergen.
\newblock A non-classical logic for information retrieval.
\newblock {\em The Computer Journal}, 25(6):481--485, 1986.

\bibitem[vR89]{RIJ89}
C.~J. van Rijsbergen.
\newblock Towards an information logic.
\newblock In N.~J. Belkin and C.~J. van Rijsbergen, editors, {\em Proceedings
  of the 12th Annual International ACM SIGIR Conference}, pages 77--86,
  Cambridge, Massachusets, 1989. ACM Press.

\end{thebibliography}
\end{document}